\documentclass[10pt,twocolumn]{article}

\usepackage{amsmath, amssymb}
\usepackage{graphics,graphicx}
\usepackage{epstopdf}
\usepackage{epsfig,comment}

\newcommand{\half}{\frac{1}{2}}

\newcommand{\lsim}{\,\raise.3ex\hbox{$<$\kern-.75em\lower1ex\hbox{$\sim$}}\,}
\newcommand{\gsim}{\,\raise.3ex\hbox{$>$\kern-.75em\lower1ex\hbox{$\sim$}}\,}

\newcommand{\fb}{\text{ fb}}

\newcommand{\TeV}{\text{ TeV}}
\newcommand{\GeV}{\text{ GeV}}
\newcommand{\MeV}{\text{ MeV}}

\begin{document}

\twocolumn[
\begin{flushright}
hep-ph/0610088\\
\end{flushright}
\begin{center}
{\huge Discovering Chiral Higgsinos at the LHC }
\vskip 0.3cm
{\normalsize
{\bf Asimina Arvanitaki}
\vskip 0.2cm
 Institute for Theoretical Physics\\
Stanford University\\
Stanford, CA 94305\\
Theory Group\\
Stanford Linear Accelerator Center\\
Menlo Park, CA 94025\\
\vskip .1in}
\end{center}

\vskip .5cm
\begin{abstract}

The concept of chirality is extended to the Minimal Supersymmetric Standard Model (MSSM) and the  $\mu$ term is forbidden by a gauged $U(1)'$ symmetry. R-parity automatically emerges after symmetry breaking, suppressing proton decay and protecting the LSP. Exotics charged under the SM pose a challenge to traditional $SU(5)$ unification, but unification is still implemented in deconstructed GUTs.  Because of the multitude of additional states to the MSSM, the $Z'$ has a large width, and the SM background, neglected in previous theoretical studies, becomes important for $Z'$ discovery. As a result, the LHC reach is reduced from 3.2 TeV, for a $Z'$ with SM decays, to 1.5 TeV, when additional decay channels are included. This model also predicts possibly long-lived colored and electroweak exotics.

\end{abstract}

\vskip 1cm
]

\section{Introduction}
\label{Sec: Introduction}

The Minimal Supersymmetric Standard Model (MSSM) \cite{Dimopoulos:1981zb} suffers from its own hierarchy problem; naturalness sets the value of the supersymmetric mass term, $\mu H_u H_d$, at the unification scale, while electroweak symmetry breaking associates $\mu$ with the vacuum expectation value (vev) of the Higgs, and demands it to be at the weak scale.

In this note, a gauged $U(1)'$ extension to the MSSM is proposed where the Higgsinos
are chiral under the new gauge symmetry, and the $\mu$ term is forbidden by gauge invariance. The fields necessary to cancel the mixed anomalies between the Standard Model gauge sector and the new $U(1)'$ form a $\mathbf{5}+\mathbf{\bar{5}}$ preserving automatic gauge coupling unification, but the colored and the electroweak parts of the fiveplets have different charges under the $U(1)'$, and a simple $SU(5)$ embedding is impossible. A solution to the split GUT multiplet problem emerges from a deconstructed orbifold GUT, $SU(5)\times SU(3) \times SU(2) \times U(1)_Y$, that is broken diagonally down to the SM gauge group.

With $U(1)'$ already forbidding  baryon- and lepton-number-violating operators, our choice of SM singlets to cancel the $U(1)'$ anomalies  generates an accidental $Z_2$ symmetry, none other than R-parity, that persists after symmetry breaking, and protects the Lightest Supersymmetric Particle (LSP).

Finally, the LHC signatures of the $Z'$ boson, the SM exotics, and the singlets are examined. The $Z'$ mass reach and forward-backward asymmetry are studied in terms of the $U(1)'$ coupling strength and the decay width, and a strong dependence of the reach on the width is found, that was previously neglected in the literature. With this choice of singlets, colored exotics are long-lived with a lifetime of 1-1000 seconds, providing distinct signatures.

\section{The $U(1)'$ Supersymmetric Standard Model}

Anomaly cancellation dictates the strategy for determining the particle spectrum. Family universality is assumed to avoid flavor changing neutral currents through the new gauge boson interactions. $U(1)'$ and $U(1)_Y$ are taken to be orthogonal, and there is no kinetic mixing that translates to millicharged particles at loop level \cite{Holdom:1985ag}.The additional particles are either singlets or vector-like under the SM, insuring anomaly cancellation within the SM gauge group. $U(1)'$ charges allow for Yukawa couplings to generate SM fermion masses. The final requirement is, of course, simplicity.

The fields chosen to cancel the mixed anomalies are a vector-like pair of down quark singlets, $D'$ and $D'{}^c$ and  a vector-like pair of lepton doublets, $L'$ and $L'{}^c$ with $U(1)'$ charges:
\begin{eqnarray}
\nonumber
L',\; L'{}^c \sim -4\\
D',\; D'{}^c\sim -6 .
\end{eqnarray}
These form a $\mathbf{5}+\mathbf{\bar{5}}$ of $SU(5)$ and automatic gauge coupling unification is preserved.   Like the $U(1)_\psi$ of $E_6$ $Z'$ models, all the Standard Model fermions have the same charge which are normalized to unity\footnote{Similar charge assignments are considered in \cite{Erler:2000wu, Morrissey:2005uz}.}. 

Notice that with these charge assignments there are no additional couplings between these exotics and the Standard Model fermions.    This forbids the $D'$ and $D'{}^c$ from acting like color-triplet Higgs which would mediate disastrous proton decay, and the $L'$ and $L'{}^c$ cannot behave like additional Higgs doublets which could give rise to flavor violating effects \cite{Hall:1993ca}. This feature also differentiates this model from $E_6$ embeddings \cite{Hewett:1988xc}.

The SM charges exclude dimension four and five baryon and lepton number violating operators:
\begin{eqnarray}
\nonumber
U^cD^cD^c, \;LLE^c,\; QD^c L \sim +3\\
QQQL,\; U^cU^c D^c E^c \sim +4
\end{eqnarray}
where $Q, U^c, D^c, L, E^c$ are the SSM chiral superfields for the Standard Model fermions.

The Higgs bosons have charge -2, which forbid the $\mu$ term
\begin{eqnarray}
H_u H_d \sim -4,
\end{eqnarray}
The Higgsino mass must now be generated after $U(1)'$ symmetry breaking
and motivates looking for TeV-scale $Z'$.  This charge assignment also forbids the final lepton number violating operator
\begin{eqnarray}
L H_u \sim -1,
\end{eqnarray}
and kinetic mixing between $L$ and $H_d$.

The charge assignments are given in Tab. \ref{Table: Charge Assignments}.
It is straight-forward to verify that this field content cancels all the mixed
anomalies and satisfies the orthogonality condition between $U(1)_Y$ and $U(1)'$.

\renewcommand{\baselinestretch}{1.3}
\begin{table}[!t]
\begin{center}
\begin {tabular}{|c||c||c|c|c|c||c|}
\hline
&N & $SU(3)$ &$SU(2)$&$U(1)_Y$&$U(1)'$& $Z_2$\\
\hline\hline
$Q$&3&$\mathbf{3}$&$\mathbf{2}$&$+\frac{1}{6}$&1&$-1$\\
\hline
$U^c$&3&$\mathbf{\bar{3}}$&$\mathbf{1}$&$-\frac{2}{3}$&1&$-1$\\
\hline
$D^c$&3&$\mathbf{\bar{3}}$&$\mathbf{1}$&$+\frac{1}{3}$&1&$-1$\\
\hline
$L$&3&$\mathbf{1}$&$\mathbf{2}$&$-\frac{1}{2}$&1&$-1$\\
\hline
$E^c$&3&$\mathbf{1}$&$\mathbf{1}$&+1&1&$-1$\\
\hline
\hline
$H_u$&1&$\mathbf{1}$&$\mathbf{2}$&$+\frac{1}{2}$&-2&+1\\
\hline
$H_d$&1&$\mathbf{1}$&$\mathbf{2}$&$-\frac{1}{2}$&-2&+1\\
\hline\hline
$D'$&1&$\mathbf{3}$&$\mathbf{1}$&$-\frac{1}{3}$&-6&+1\\
\hline
$D'^c$&1&$\mathbf{\bar{3}}$&$\mathbf{1}$&$+\frac{1}{3}$&-6&+1\\
\hline\hline
$L'$&1&$\mathbf{1}$&$\mathbf{2}$&$-\frac{1}{2}$&-4&+1\\
\hline
$L'^c$&1&$\mathbf{1}$&$\mathbf{2}$&$+\frac{1}{2}$&-4&+1\\
\hline
\end{tabular}
\renewcommand{\baselinestretch}{1.0}
\caption{\label{Table: Charge Assignments} Charge assignments for the $U(1)'$ SSM and the exotics.}
\end{center}
\end{table}
\renewcommand{\baselinestretch}{1.0}

\renewcommand{\baselinestretch}{1.3}
\begin{table}[!t]
\begin{center}
\begin {tabular}{|c||c||c|c|c|c||c|}
\hline
&N & $SU(3)$ &$SU(2)$&$U(1)_Y$&$U(1)'$& $Z_2$\\
\hline
\hline
$N_4$&2&1&1&0&4&+1\\
\hline
$N_8$&2&1&1&0&8&+1\\
\hline
$N_{12}$&1&1&1&0&12&+1\\
\hline\hline
$S_{-2}$&1&1&1&0&-2&+1\\
\hline
$S_{-6}$&7&1&1&0&-6&+1\\
\hline
$S_3$&9&1&1&0&3&$-1$\\
\hline
\end{tabular}
\renewcommand{\baselinestretch}{1.0}
\caption{ \label{Table: Charge Assignments Singlets} Charge assignments for the singlets.}
\end{center}
\end{table}
\renewcommand{\baselinestretch}{1.0}

In order for the Higgsinos and the Standard Model exotics to acquire masses at the 
weak scale singlets need to be added so that their vevs
give rise to TeV scale masses through renormalizable interactions
\begin{eqnarray}
\lambda_H N_4 H_u H_d + \lambda_{L'} N_8 L' L'{}^c + \lambda_{D'} N_{12} D' D'{}^c .
\end{eqnarray}

Finally singlets are added to cancel the $U(1)'^3$ and mixed gravitational $U(1)'$anomalies. These singlets have to be chiral in order not to reintroduce the $\mu$ problem with mass terms in the superpotential. Several interactions between the scalars are needed to drive  all the vevs at roughly the same scale;  supersymmetry breaking should trigger the $U(1)'$ breaking
and any significant ratio between the vevs would either lead to fine-tuning
of the Higgs potential or having too light exotics.    This is easy to
arrange if there are many Yukawa couplings between the different fields.
In Tab. \ref{Table: Charge Assignments Singlets} a choice for these
singlets is presented that satisfies the requisite features and has no flat directions in the superpotential insuring stability after symmetry breaking.  It should be emphasized that
this is  a choice of singlets that works well for phenomenology and 
most aspects of the phenomenology follow through.

The superpotential is given by
\begin{eqnarray}
\nonumber
W&=&\kappa_H  N_4 S_{-2}^2 + \kappa_D  N_{12}S_{-6}^2 + \kappa_L N_8 S_{-6} S_{-2}\\
&&+ h S_{-6} S_{3}^2 +  y S_3 L L'{}^c
\end{eqnarray}
The last term gives rise to small lepton flavor violating effects the consequences  of which are discussed in Sec. \ref{Sec: Particle Physics Limits}.
The color exotics can not decay until dimension 5 operators:
\begin{eqnarray}
N_8 S_3 D^c D',\; S_4 U^c D^c D'{}^c,\; S_4  Q L D^c.
\end{eqnarray}
This results in a lifetime of 10 seconds and the limits are discussed in Sec. 
\ref{Sec: Cosmological Limits}.  The latter two operators violate baryon number
and lepton number respectively and lead to dimension 6 proton decay.

\subsection{Symmetry Breaking}

As advertized, the $\mu$ term is generated with symmetry breaking:
\begin{eqnarray}
\mu \equiv \lambda_H \langle N_4 \rangle,
\end{eqnarray}
in addition to the $B_{\mu}$ term, $B_{\mu} h_u h_d$, :
\begin{eqnarray}
B_{\mu} \equiv \lambda_H \kappa_H \langle S_{-2} \rangle^2,
\end{eqnarray}
and they are both  automatically at the right scale, since the $U(1)'$ symmetry breaking is associated with SUSY breaking. As a result, the Higgs phenomenology is expected to be similar to the NMSSM one. At the same time, all the singlets and exotics get masses at the 100 - 1000 GeV range, when $N_4, ~ N_8, ~ N_{12}$ and $S_{-6}$ get vevs. This is insured by the Yukawa couplings, the potential and the A terms; D terms contribute to the negative mass terms, while A terms and F terms, because of the Yukawa couplings, provide linear in the fields terms that drive the vevs away from zero. 

The value of the Higgs quartic is increased from a small contribution from the F-terms, because of the $N_4 H_u H_d$ term of the superpotential.
\begin{eqnarray}
\lambda=\frac{g_1^2+g_2^2}{8}\cos^2(2\beta)+\frac{\lambda_H^2}{4}\sin^2(2\beta),
\end{eqnarray}
increasing slightly the Higgs mass and reducing the MSSM fine-tuning \cite{Dermisek:2005ar}.

Due to an accidental $Z_3$ symmetry of the renormalizable superpotential, there are domain walls created with symmetry breaking. But the $Z_3$ symmetry is violated by GUT-scale-suppressed dimension-five operators, resulting in a pressure difference of $\sim \frac{M^5_{U(1)'}}{M_{GUT}}$ between the different vacua that is enough to push the domain walls outside our horizon long before BBN \cite{Abel:1995wk}. 

\subsubsection*{R-parity}

In addition to all lepton and baryon number violating operators being forbidden by gauge invariance, there is an accidental $Z_2$  symmetry in the superpotential (Tab. \ref{Table: Charge Assignments} and \ref{Table: Charge Assignments Singlets}); this is nothing else but R-parity, the Lightest Supersymmetric Particle (LSP) is stable and there is still a viable Dark Matter (DM) candidate. The neutralino sector is extended by the singlets of charge +1, without affecting the nature of the MSSM bino LSP, because of  precision electroweak and cosmological constraints (Sec. \ref{Sec: Cosmological Limits} and \ref{Sec: Particle Physics Limits}). 

R-parity is preserved provided that $S_3$ does not acquire a vev. This is improbable since $S_3$ is a point of increased symmetry with just one Yukawa coupling and another approximate $Z_2$ symmetry with $S_3, ~ L'$ and $L'^c$ having -1 charge under it. This symmetry is broken at the non-renormalizable level, rendering the scalar part of $S_3$ long-lived, with a lifetime constrained by BBN to be less than 1 second (Sec. \ref{Sec: Cosmological Limits}).

\section{Preserving Gauge Coupling Unification}
\label{Sec: GCU}

From the requirement of simplicity alone we find adding a {\boldmath{$5 \oplus \bar{5}$}} cancels all mixed anomalies, and the SM couplings still unify at the GUT scale. But, what is the group they unify to? $D'$ and $L'^c$ as well as  $D'^c$ and $L'$ form a {\boldmath{$5$}} and a {\boldmath{$\bar{5}$}}, respectively, but these fiveplets do not have universal $U(1)'$charges, and an $SU(5)$ embedding does not commute with $U(1)'$. We could assume that $D'$, $D'^c$, $L'$ and $L'^c$ come from four different fundamentals and anti-fundamentals of $SU(5)$, but doublet-triplet splitting is not possible without breaking $U(1)'$ at the GUT scale or having additional split multiplets . The above particle spectrum is not consistent with an $E_6$ embedding \cite{Hewett:1988xc} either, which, besides a very different particle spectrum that includes right-handed neutrinos - not a necessity in this model-, predicts again universal charges for the fiveplets. In fact, any $G \times U(1)'$ embedding is excluded, because it is impossible to produce split GUT multiplets without breaking $U(1)'$ at the GUT scale.


An answer to the question of unification is found in $SU(5) \times SU(3) \times SU(2) \times U(1)_Y \times U(1)'$, a group inspired by orbifold deconstruction \cite{Weiner:2001pv, Csaki:2001qm}. This group is broken diagonally down to $SU(3)\times SU(2) \times U(1)_Y$ with the high scale SM couplings given by:
\begin{eqnarray}
\frac{1}{g_{i}^2}=\frac{1}{g_{5_{GUT}}^2}+\frac{1}{g_{i_{GUT}}^2}
\end{eqnarray}
If $SU(3)\times SU(2) \times U(1)_Y$ is strong at the symmetry breaking scale, then the above relations translate to:
\begin{eqnarray}
{g_i} \approx {g_5}_{GUT} \text{, i=1, 2, 3}
\end{eqnarray}
and the SM gauge dynamics are determined by one coupling. This condition also insures that the hypercharge normalization does not change, another condition for SSM unification. Quarks and leptons are the usual {\boldmath{$10 \oplus \bar{5}$}} of $SU(5)_{GUT}$, while the Higgs doublets and the exotics are doublets and triplets of $SU(3)\times SU(2) \times U(1)_Y$. 

Yukawa couplings and the renormalizable low-energy couplings involving quarks, leptons and the exotics appear at the GUT scale as non-renormalizable operators. For example, the Yukawas are generated from:

\begin{eqnarray}
\frac{{\bf{10 \ 10}} \ \Sigma_1 \  H_u}{M} ~ ~\text{and} ~ ~
\frac{{\bf{10 \ \bar{5} }}\ \Sigma_2  \ H_d}{M},
\end{eqnarray}

where $\Sigma_1$, a $(5,1,2, -\frac{1}{2},0)$, and $\Sigma_2$, a $(\bar{5},1,2, \frac{1}{2},0)$,  get vevs with symmetry breaking.

\section{Experimental Limits}

\subsection{Cosmological Limits}
\label{Sec: Cosmological Limits}

The first bounds come from cosmology; this model generically has long lived states, possibly charged. Depending on their lifetime, their decay affects the primordial abundance of light elements during BBN \cite{Kawasaki:2004qu}, discrupts the cosmic microwave background, or contributes to the diffuse gamma ray spectrum \cite{Kribs:1996ac, Hu:1993gc}. If their lifetime is larger than the age of the universe and they are charged, they form heavy elements \cite{Smith:1982qu, Hemmick:1989ns}. With this choice of singlets,  the fermion colored exotics are long-lived and they decay through the same operators that mediate proton decay to a quark and a squark:
\begin{eqnarray}
\nonumber
\hspace{-0.2in} \Gamma_{D'} \hspace{-0.1in}&\approx& \frac{1}{16 \pi} \frac{ \langle S_4 \rangle^2}{M_{GUT}^2}m_{D'}\\
\hspace{-0.2in}& \sim&\hspace{-0.1in}(1000 \text{ sec})^{-1} 
\bigg(\frac{\langle S_{4} \rangle}{100 \GeV}\bigg)^2\bigg(\frac{ m_{D'} }{100 \GeV}\bigg).
\end{eqnarray}
If the fermion $D'$ is lighter than the squark, these exotics are excluded from heavy element searches. The decay of $D'$ relics affects the $^3He$ abundance, but they annihilate significantly through the strong interactions without placing any constraints on the relevant mass scales.

As already discussed, there is an approximate $Z_2$ symmetry involving the singlets with negative R-charge and the lepton-like exotics. This symmetry is broken by dimension five operators and there is another long-lived state with the fermion $S_3$ and $L'$ as candidates. If $S_3$ is lighter than  $L'$, it decays through a dimension five operator
$S_{-2} S_3LH_u$:
\begin{eqnarray}
\Gamma_{S_3}\sim (1000 \text{sec})^{-1} \bigg( \frac{\langle S_{-2} \rangle}{100  \GeV}\bigg)^2\bigg(\frac{m_{S_3}}{100  \GeV}\bigg) 
\end{eqnarray}
If the $Z'$ boson mass and the $g'$ coupling saturate precision electroweak constraints in the relic abundance calculation, $^6Li$ sets the upper limit on the lifetime. Assuming all decay products are hadronic, this value is $\sim 1 \ \text{sec}$ and the corresponding bound on $y \langle S_{-2} \rangle$ and $m_{S_3}$about $500-1000 \GeV$.

 If $L'$ is lighter than $S_3$, it decays through a  dimension seven operator
 $ S_4 S_{-2} L' L E$ or $S_4 S_{-2} L' Q D^c$:
\begin{eqnarray}
\nonumber &&\Gamma_{L'} \sim \frac{ \langle S_4 \rangle^2 \langle S_{-2} \rangle^2}{M_{GUT}^6}m_{L'}^3 \sim (10^{60} \text{ sec})^{-1}
\end{eqnarray}
It is the neutral $L'$ that is long-lived, since electroweak symmetry breaking produces a splitting of $\sim 200-350  \MeV$ between the charged and the neutral component of the doublet with the charged component being heavier\cite{Thomas:1998wy}.  This is excluded from direct DM searches, because neutral Dirac fermions have vector couplings to the Z boson and they scatter coherently with the nuclei of the detector material, resulting in a large interaction cross-section\cite{Akerib:2005kh, Goodman:1984dc}.

\subsection{Particle Physics Limits}
\label{Sec: Particle Physics Limits}

Precision electroweak data put further limits on the vev of the fields giving mass to the $Z' $ boson \cite{Barbieri:2000gf}: 
\begin{eqnarray}
\frac{Q^2_{N} \langle N_{12}\rangle^2}{Q_H Q_{SM}} < (9.2 \TeV)^2,
\end{eqnarray}
for a Higgs mass of 115 GeV.  This translates to a lower bound of 
\begin{eqnarray}
\langle N_{12}\rangle > 1.1  \TeV .
\end{eqnarray}
If the gauge coupling for the $U(1)'$ is strongly coupled at the GUT scale, then
this corresponds to a $650 \GeV$ limit on the mass of the $Z'$.

There is an additional source of lepton flavor violation in this mode from the 
couplings of Standard Model leptons to the lepton exotics:
\begin{eqnarray}
y_{l}{S_3} L_lL'^c.
\end{eqnarray}
This leads to a contribution to $\mu \rightarrow e \gamma $, and the corresponding branching fraction is:
\begin{eqnarray}
 \frac{12\pi y_{\mu}^2y_e^2\alpha_{EM}}{M^4 G_F^2} \sim 10y_{\mu}^2y_e^2\Big( \frac{100 \GeV}{M} \Big) ^4,
\end{eqnarray}
where  $M$ is the scale associated with the mass of $S_3$ and $L'$.   If we assume
minimal flavor violation for the couplings, $y_l \sim \sqrt{{y_{SM}}_l}$, then
the current experimental bound is a branching fraction of $10^{-12}$ \cite{Eidelman:2004wy}, which
translates to  a  limit:
\begin{eqnarray}
M \gsim 1 \TeV.
\end{eqnarray}

The colored and electroweak exotics have no Yukawa couplings to the Higgs and no mixing with SM particles at the renormalizable level, since $S_3$ does not get a vev. There is mixing between $L'$ and $H_u$, as well as $L'^c$ and $H_d$,  through dimension five operators, but it is suppressed by $M_{GUT}$. The bound on the masses thus comes from direct searches alone and should be set at around $200 \GeV$ for $D'$ and $100 \GeV$ for $L'$, a number extracted from other searches for electroweak doublets and color triplets \cite{Eidelman:2004wy}.

\section{LHC Signatures}

\subsection{$\mathbf{Z'}$ Physics}
Adding a gauged $U(1)'$ symmetry to the SM has been studied extensively for both the Tevatron and the LHC, especially in the framework of $E_6$ inspired models. The main discovery channel is the detection of a resonance in the invariant mass distribution of events with two leptons of the same flavor in  the final state. There are more channels available for quarks in the final state, but the signal is obscured by QCD background. The $Z'$ reach of a collider depends on both the value of the $Z'$ coupling and the SM charges, as well as the width of the resonance. The importance of the latter has been ignored in the literature by  assuming that the $Z'$ decays only to SM particles. But the requirement of weak scale SUSY alone introduces new states to which the $Z'$ decays, increasing its width. This results in a broader $Z'$ resonance which means increased background and reduced $Z'$ production cross-section at the same time. The width of the $Z'$ in terms of the couplings and the charges of the states it decays to is, assuming these states are massless:
\begin{displaymath}
\displaystyle
\Gamma_{Z'}=\frac{g^2M_{Z'}}{24\pi}(\sum_{\text{\tiny{fermions}}}q^2_f+\half \sum_{\text{\tiny{scalars}}}q^2_s)\\
\equiv \frac{g^2M_{Z'}}{24\pi}\beta
\end{displaymath}
$\beta$ becomes proportional to the $\beta$-function of the $g$ coupling when all states that are charged under the $Z'$ are available for decay. In that case, if  $U(1)'$ does not become strong until the GUT scale, g at low energies is proportional to $\frac{1}{\sqrt{\beta}}$; a small coupling is related to a large width.

\begin{figure}[!t]
\begin{center}
\includegraphics[width=3in]{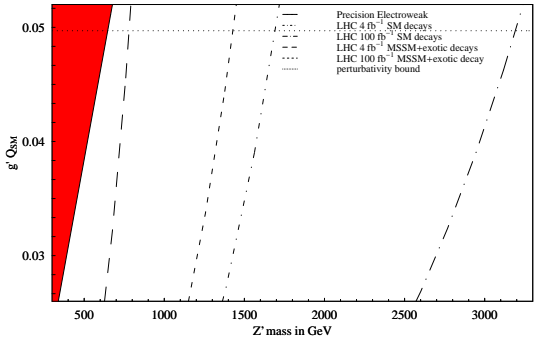} 
\caption{ \label{Fig:Z' bounds}  The LHC projected $Z'$ mass reach for 4 fb$^{-1}$ and 100 fb$^{-1}$ for a $Z'$ with SM decays and MSSM+exotic decays as a function of the coupling. The red area is the currently excluded by precision electroweak region.}
\end{center}
\end{figure}

\begin{figure}[!t]
\begin{center}
\includegraphics[width=3.0in]{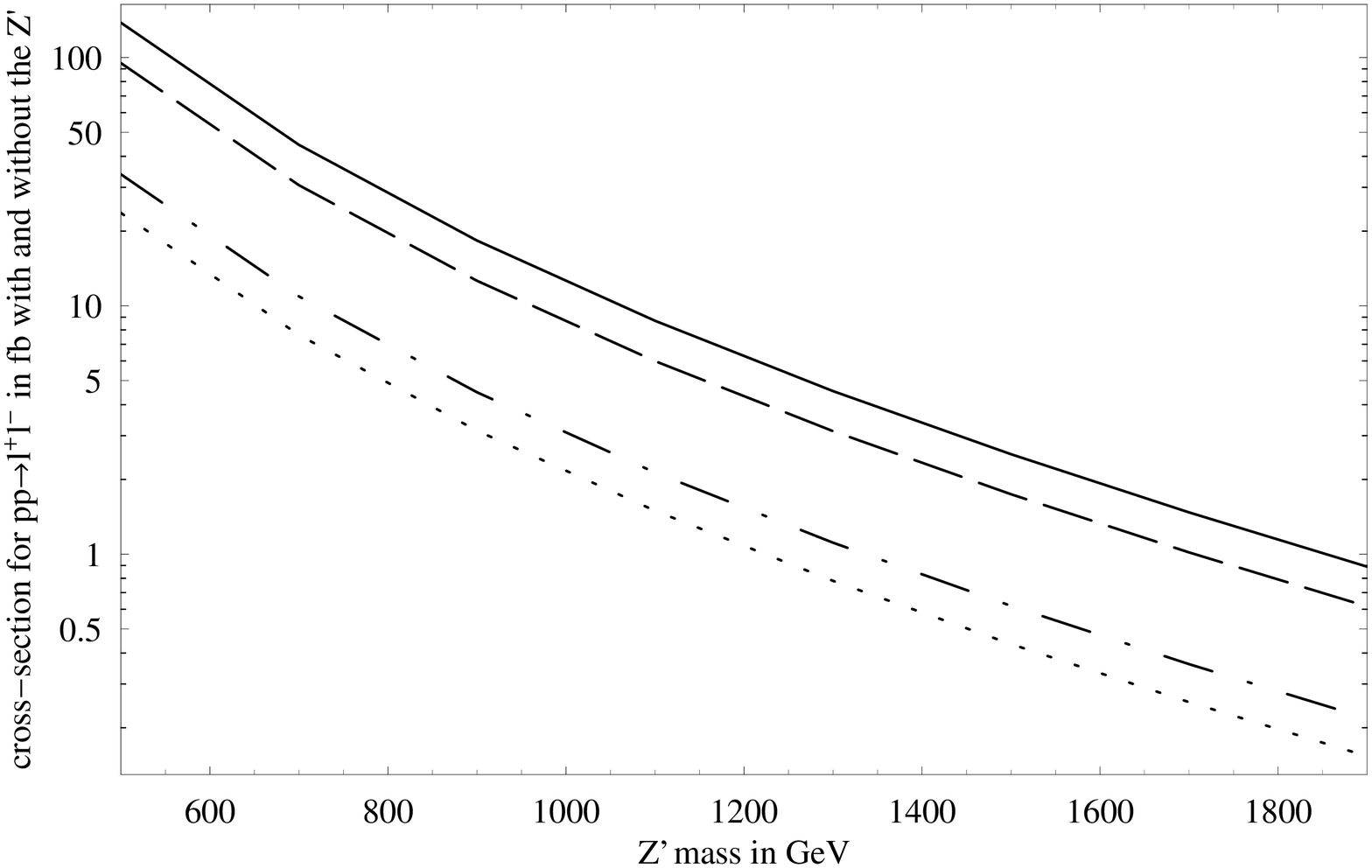}
\caption{ \label{Fig:Z' production} The production cross-section of $e^+e^-$ at the LHC in a $2 \Gamma$ window around the $Z'$ resonance as a function of the $Z'$ mass. The full line is for $g' Q_{SM}=0.05$, the perturbativity bound, while the dashed line is the corresponding background. Similarly, the dot-dashed and the dashed lines are production cross-sections with and without the $Z'$, respectively, for $g' Q_{SM}=0.025$.}
\end{center}
\end{figure}

The signal to background ratio is:
\begin{eqnarray}
\frac{\text{S(ignal)}}{\text{B(ackground)}} = \text{const}_1\frac{\frac{g^4}{\Gamma_{Z'}}}{\Gamma_{Z'}} = \frac{\text{const}_1}{\beta^2},
\end{eqnarray}
and the luminosity necessary to establish $5\sigma$ discovery is:
\begin{eqnarray}
\text{L}=\frac{25(\sigma_B+\sigma_S)}{\sigma_S^2}\sim \frac{( \frac{\beta^3}{\text{const}_1}+\beta)}{g^2}
\end{eqnarray}
When the width is small, just as in the case of a $Z'$ with SM decays, L becomes proportional to the width:
\begin{eqnarray}
\text{L}\sim\frac{\beta}{g^2},
\end{eqnarray}
while in the case where the $Z'$ has a many decay channels:
\begin{eqnarray}
\text{L}\sim\frac{\beta^3}{g^2},
\end{eqnarray}
which makes clear that the decays of the $Z'$ to non SM states greatly affect the discovery reach. This is also seen in Fig. \ref{Fig:Z' bounds}, where  we present the LHC reach for 4 and 100 $\fb^{-1}$ of integrated luminosity as a function of the coupling strength for a $Z'$ with just SM decays and a $Z'$ that decays to all MSSM and fermion singlet states. This is equivalent to a width to mass ratio of about $2.3\%$, an order of magnitude larger from that in the case of just SM decays ($\frac{\Gamma}{M}\sim0.17\%$).  The cross-sections are evaluated with CalcHEP \cite{Pukhov:2004ca} and we have set a 20 GeV cut in the transverse momentum.  The  detection limit at the LHC for a  $Z'$ with additional to the SM decay channels is reduced  to  1.5 TeV, compared to $\sim 3.25 ~ \text{TeV}$ for one with SM decays, for an integrated luminosity of 100 fb$^{-1}$, see also Table  \ref{Table: Reach}. Furthermore, in Fig. \ref{Fig:Z' production} we present total dilepton production cross-section at the LHC with and without the $Z'$ for two different values of the coupling, as a simple verification of the previous equations. The total cross-section goes down by a factor of 4,  the square of the ratio of the couplings, while the signal to background ratio does not change.

\renewcommand{\baselinestretch}{1.3}
\begin{table}[!t]
\begin{center}
\begin {tabular}{|c||c|c|}
\hline
Model&Width/Mass& Reach (GeV)\\
\hline
\hline
Just SM decays&$0.17\%$ &3200\\
\hline
In this paper&$2.3\%$&1500\\
\hline
\end{tabular}
\renewcommand{\baselinestretch}{1.0}
\caption{ \label{Table: Reach} LHC $Z'$ mass reach for two different cases of $U(1)_{\psi}$ at 100 $\text{fb}^{-1}$.}
\end{center}
\end{table}
\renewcommand{\baselinestretch}{1.0}

\begin{figure}[!t]
\begin{center}
\includegraphics[width=3in]{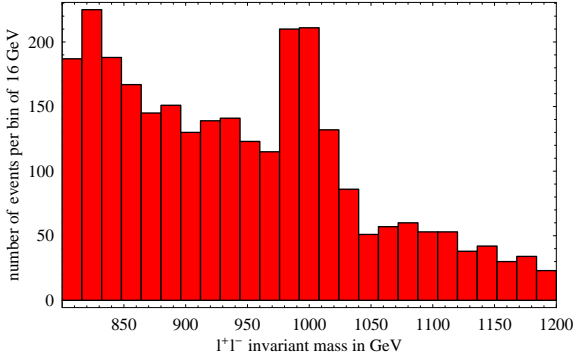}
\caption{ \label{Fig:Z' resonance} The invariant mass distribution of both $e^+e^-$ and $\mu^+\mu^-$ around the $Z'$ resonance. The mass of the $Z'$ is set at 1000 GeV and the width at 23 GeV. }
\end{center}
\end{figure}

After discovery, the next step is determining  the $Z'$ couplings. The first information about the value  of $g' Q_{SM}$ and exotic states is contained in the resonant production cross-section and the width, which may be the only evidence we acquire for particle states beyond the MSSM. In Table \ref{Table: Fitted Parameters}, we present the measured values when the $Z'$ mass is 1 TeV. The integrated luminosity at the LHC is 100 $\text{fb}^{-1}$. Events have been generated using the Harvard blackbox\cite{Thaler:2004}, which integrates Pythia \cite{Sjostrand:2003wg} and PGS\cite{Conway:2004}, while providing a simple data manipulation mathematica program, the Chameleon. The determination of the cross-sections has been performed by simple event counting in a $3\Gamma$ window around the resonance, while the mass and the width have been determined by fitting the peak as a Breit-Wigner resonance. The background has been subtracted by generating dilepton events that do not include the $Z'$. For a 1 TeV $Z'$ with 100 $\text{fb}^{-1}$ of events at the LHC the mass is determined with an accuracy better than $1\%$, while for the other parameters the corresponding error is $10-30\%$, when the discovery luminosity for a 1 TeV $Z'$ is just a factor of four smaller, $\sim 25~\fb^{-1}$.  In the cross-section and the ratio $\frac{\sigma_{\mu^+\mu^-}}{\sigma_{e^+e^-}}$ error estimate we have included an additional $10\%$ error to account for systematics, e.g. luminosity, parton distribution functions (PDFs) and detector efficiency uncertainties.
\renewcommand{\baselinestretch}{1.3}
\begin{table}[!t]
\begin{center}
\begin {tabular}{|c||c|c|}
\hline
Parameter & Exp. Value & Mod. Value\\
\hline
\hline
Mass&$992.4\pm 2.3$ GeV&1000 GeV\\
\hline
Width&$29\pm 7$ GeV &23 GeV\\
\hline
$pp \stackrel{Z'}{ \longrightarrow}  e^+ e^-$&$2.04 \pm 0.25$ fb&2.33 fb \\
\hline
$pp \stackrel{Z'}{ \longrightarrow} \mu^+ \mu^-$&$2.00 \pm 0.25$ fb& 2.33 fb\\
\hline
 $\frac{\sigma_{\mu^+\mu^-}}{\sigma_{e^+e^-}}$&$0.98 \pm 0.17$& 1 \\
\hline
\end{tabular}
\renewcommand{\baselinestretch}{1.0}
\caption{ \label{Table: Fitted Parameters} Determination of $Z'$ mass, width, and cross-section to leptons at 100 $\text{fb}^{-1}$.We have estimated from the generated data the electron efficiency to be $95\%$ and the muon efficiency $97\%$.}
\end{center}
\end{table}
\renewcommand{\baselinestretch}{1.0}

\subsubsection*{Forward-Backward Asymmetry}
\begin{figure}[!t]
\begin{center}
\includegraphics[width=3.0in]{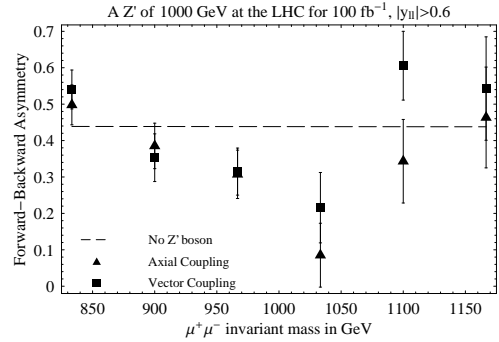}
\caption{ \label{Fig:FB asymmetry} The forward-backward asymmetry of $\mu^+\mu^-$ through the $Z'$ at the LHC for two a vector and an axial symmetry as a function of the invariant mass. The bin size is 33.3 GeV and there is a transverse momentum cut of 20 GeV. }
\end{center}
\end{figure}

The  observable, though, that helps reveal the axial character of the symmetry is the forward-backward asymmetry of the two leptons produced through the photon, the $Z$, and the $Z'$:
\begin{eqnarray}
\text{A}_{\text{FB}}=\frac{\int^1_0 \sigma(x)dx-\int^0_{-1} \sigma(x)dx}{\int^1_0 \sigma(x)dx+\int^0_{-1} \sigma(x)dx},
\end{eqnarray}
where $x \equiv \cos \theta^*$, and $\theta^*$ is the angle between the lepton and the quark in the lepton-pair center-of-mass frame. This measurement is possible even in a pp collider because the information of the quark's direction is preserved for leptons produced in the forward direction ($|\eta|>0.6$)\cite{Dittmar:1996my} in the boost necessary to go to the center of mass of the dileptons. Interference between the SM gauge bosons is also essential for this effect, as $U(1)'$ has no vector couplings and has to interfere with the vector couplings of SM gauge bosons, unlike the SM Z, which has both axial and vector couplings. Fig.  \ref{Fig:FB asymmetry} presents the expected forward-backward asymmetry for our axial $U(1)$ in comparison with a vector $U(1)$ with exactly the same interaction strength for a $Z'$ of 1 TeV mass at 100 $\mbox{fb}^{-1}$ at the LHC. Going through the $Z'$ resonance, a few $\sigma$ deviation from the SM value and just a 2-3 $\sigma$ difference from the corresponding values of the axial U(1) are found. Both these statistical deviations are affected by  the U(1) coupling strength and the width of the $Z'$. The value of the on-peak FB asymmetry is, in terms of g and $\beta$:
\begin{eqnarray}
\text{A}_{\text{FB}}=\text{const}_2 \frac{ (g^2 +  \text{const}_3) \beta^2}{(1+\frac{\beta^2}{\text{const}_1})},
\end{eqnarray}
while the corresponding error is:
\begin{eqnarray}
\nonumber
\Delta \text{A}_{\text{\tiny{FB}}}= &\frac{1}{\sqrt{\text{L}}} \sqrt{\frac{(1-{\text{A}^2}_{\text{\tiny{FB}}})}{(\sigma_S+\sigma_B)}} \\
&\sim\frac{1}{\sqrt{\text{L}}} \sqrt{\frac{(1-{\text{A}^2}_{\text{\tiny{FB}}})}{\frac{g^2}{\beta} (1+\frac{\beta^2}{\text{const}_1})}}.
\end{eqnarray}

The asymptotic value of the standard deviations away from the SM value for a given luminosity is, in the limit of large $\beta$ and small g:
\begin{eqnarray}
&&\frac{|\text{A}_{\text{\tiny{FB}}}-\text{A}_{\text{\tiny{FB-SM}}}|}{\Delta \text{A}_{\text{\tiny{FB}}}}\nonumber \\ 
&&\sim \sqrt{\text{L}g^2} \arrowvert \text{const}_2~ g^2 \sqrt{\beta} - \frac{\text{A}_{\text{\tiny{FB-SM}}}}{\sqrt{\beta^3}} \arrowvert
\end{eqnarray}
This $\beta$ dependence translates to a decreasing deviation from the SM value with increasing width. $\beta$ is related to the running of g, a small g follows a large width, and, since g is small compared to SM couplings, the second term is important.

In the example presented in Fig. \ref{Fig:FB asymmetry}, a luminosity of 100 $\text{fb}^{-1}$ is a factor of four larger than the luminosity necessary for discovery of the $Z'$, and this translates to 2-3$\sigma$ deviation from the SM value for the FB asymmetry when discovery is established. As a result, the effects of the $Z'$ resonance in combination with the excess of events in the invariant mass distribution help establish discovery with fewer events, increasing the LHC mass reach.

\subsubsection*{Increasing Sparticle Reach}
The $Z'$ boson is a distinctive signature on its own, but it is also a new channel of beyond the SM particle production. It increases the total production cross-section for a given MSSM state, and, being heavy, results in events with higher transverse momentum. For the case of left-handed sleptons, we examine at the transverse mass, $M_T$, distribution of events that have two leptons of the same flavor and missing energy in the final state, where $M_T$ is defined as:
\begin{eqnarray}
M^2_T=(P^\mu_{\ell}+P^\mu_{\bar{\ell}}+\text{missing} ~P^\mu_{T})^2
\end{eqnarray}
CompHEP \cite{Boos:2004kh}and the CompHEP-PYTHIA interface \cite{Belyaev:2000wn} are used to generate the events. We set a 125 GeV missing energy cut on the events along with a 40 GeV veto on jets, and a 50 GeV $p_T$ cut. A cut in the azimuthal angle between the two leptons, $\Delta \phi_{e^+ e^-} < 160^o$, is imposed, and the diboson and $t\bar{t}$ SM backgrounds are included. The analysis is performed for a 1 TeV $Z'$ with a coupling that saturates the perturbativity bound. In the case where the $Z'$ can decay to all MSSM and fermion singlet states we find no significant increase in the reach.   When only MSSM states are available for decay, the reach is increased by $\sim 10\%$ compared to the left-handed slepton reach in the MSSM \cite{Dittmar:1998rb}, which in terms of numbers is 450 GeV at $100~\fb^{-1}$, or in other words  a $30\%$ decrease in the luminosity necessary to establish $5\sigma$ discovery in the absence of a $Z'$.  The reach also depends on possible decay modes of the sleptons. Here, the sleptons are assumed to decay only to the LSP of 100 GeV mass. If this mass is increased or there are other neutralino or chargino states below the slepton mass the reach is reduced. 

\subsection{The exotics}

The exotics complement the $Z'$ discovery. 

\subsubsection*{Colored Exotics}
In the absence of a singlet with charge +5, colored exotics have a lifetime of 1 to 1000 seconds. The scalar colored exotics are typically heavier, since their soft mass terms add to the mass from the $U(1)'$ symmetry breaking. They decay to the exotic fermion state and either a bino, gluino or $Z'$-ino, depending on the amount of phase space available for the decay.   The subsequent cascade decays provide a detectable signal.

It is the fermions of this supermultiplet that provide striking signatures. They hadronize shortly after they are produced to an isodoublet R-hadron. As they travel through the detector, the ones with $\frac{v}{c}>\frac{1}{3}$ leave stiff tracks and timing measurements in the muon chamber give a mass reach of 1.3 TeV \cite{Kraan:2005ji}. 

\begin{figure}[!t]
\begin{center}
\includegraphics[width=3.0in]{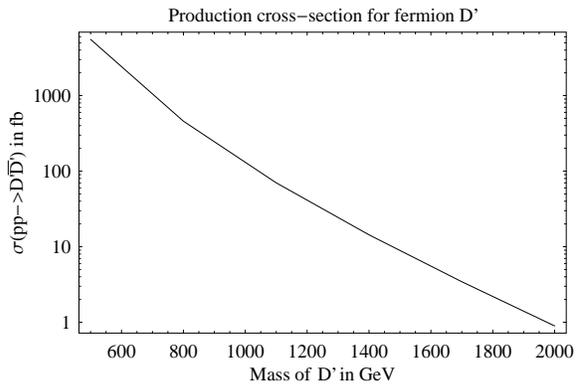} 
\caption{ \label{Fig:D' production} The production cross-section of $D'$ fermions at the LHC as a function of their mass}
\end{center}
\end{figure}

When the slow ones, $\frac{v}{c}<\frac{1}{3}$,  enter the dense material of the calorimeter, strong interactions make the R-hadron oscillate between charged and neutral states. While they are charged, electromagnetic losses slow them down and stop them.  After stopping they wait inside the calorimeter until they decay, giving
striking signatures much like long-lived gluinos in split supersymmetry \cite{Arvanitaki:2005nq}. These decays are uncorrelated with
any activity in the detector and in particular they appear as events with large missing energy with no
tracks pointing to the interaction region.  

Measuring the lifetime of these exotics gives us much less information than the long-lived gluinos of split
SUSY since they decay through GUT-supressed dimension 5 operators.

In addition, there are monojets associated with the production of these long-lived states, giving a reach of 1.1 TeV at the LHC for 100 fb$^{-1}$
\cite{Hewett:2004nw}.

\subsubsection*{Electroweak Exotics}

On the other hand, the lepton-like exotics are out of reach; BBN constraints set the mass of $S_3$ larger than 500-1000 GeV, while direct searches require excluded stable neutral $L'$ states, and they have to decay at the renormalizable level to a singlet and a lepton. At masses larger than 500 GeV there are very few production events, insufficient to give a signal above the SM background.

With other charge assignments and masses these state may have short lifetimes or even be quasi-stable
like the colored exotics are.  Their stopping is roughly similar to a $\tilde{\tau}$ NLSP and their stopping
was considered in   \cite{Feng:2004yi}.

\subsection{Singlets}
Finally, there are the exotics that are singlets under the SM. The fermion part of the lightest singlet with -1 R-charge is quasi-stable and decays outside the detector. The rest of the fermion states will cascade decay down to the lightest. The scalar part is heavier and will decay to a neutralino and the corresponding fermion state. The fermion component of the singlets with +1 R-charge extend the neutralino sector and once produced, they cascade decay down to the LSP. The scalar part again decays to a neutralino and the corresponding fermion singlet. These cascade decays result in similar to the MSSM multi-jet and multi-lepton signatures, so the search strategy is similar to the one for neutralinos, but it is impossible for the LHC to decipher the full structure of the singlet sector.

\section{Conclusion}

Protecting the $\mu$ term with a gauged $U(1)'$ symmetry naturally gives rise to R-parity, that insures proton stability and a viable Dark Matter candidate. This is an old idea; Weinberg first pointed out that the MSSM problems with proton stability can be solved by adding a gauged U(1) \cite{Weinberg:1981wj}, and a multitude of papers inspired by $E_6$ models followed reference \cite{Hewett:1988xc}.  Despite similarities to $U(1)_{\psi}$ of $E_6$, our $U(1)'$ distances itself from the wide class of $E_6$ models with different charge assignments for the exotics and the singlets. These assignments automatically forbid dangerous couplings to SM particles that mediate proton decay and flavor violation without the need of additional symmetries, unlike the case of $E_6$ embeddings. The exotics's charge assignments also create an unsolvable doublet-triplet splitting problem for usual unification, but the latter can still be preserved in a deconstructed GUT group embedding. 


A particle spectrum similar to the one presented here is found in \cite{Erler:2000wu} and \cite{Morrissey:2005uz}. In \cite{Erler:2000wu}, though, the electroweak exotics have the same quantum numbers as the Higgs doublets, while in \cite{Morrissey:2005uz}  the potential of the singlets is not fully explored, there are singlets that are vector-like under $U(1)'$, and an R-parity is imposed.

The additional states to the MSSM further affect the phenomenology of the model through the $Z'$ width. The mass reach for the $Z'$ is decreased as the branching fraction to non-SM states increases; the luminosity necessary for discovery varies as $\Gamma^3$  for large $\frac{\Gamma}{\text{M}}$, a stronger dependence than that in the case of a narrow resonance, where the background is negligible, and $\text{L}_{\text{\tiny{discovery}}}\sim \Gamma$. The LHC reach  from 3.2 TeV for a $Z'$ with SM decays drops to 1.5 TeV, when additional decay channels are considered. These states also determine the accuracy at which the different $Z'$ observables, such as the forward-backward asymmetry, can be measured.

The $Z'$ itself provides an additional channel for sparticle production and can improve the LHC reach providing events with higher transverse momentum, away from the SM background. The picture is completed by the singlets and the exotics; even though the full structure of the singlet sector cannot be resolved, there are colored and electroweak exotics that are possibly long-lived, and make this model an exciting prospect for the LHC.

\section*{Acknowledgements}

First, I would like to express my gratitude to Jay Wacker, who participated in some stages of this work, for all his help and guidance. I would also like to thank Savas Dimopoulos for pointing out deconstructed GUT models and for many useful discussions. Special thanks to the Harvard blackbox team, in particular Tom Hartman for his help with the comphep-pythia interface, and Can Killic for spending a lot of time debugging code with me. Many thanks to Ignatios Antoniadis for motivating work on gauged extensions of the MSSM, and Tom Rizzo for useful discussions on $Z'$ phenomenology. I would finally like to thank the Galileo Galilei Institute in Florence and Johns Hopkins University for their hospitality during completion of this work.

\end{document}